\documentclass[12pt]{article}
\usepackage{amsthm,latexsym,multibox,amssymb}
\usepackage{graphicx,lscape,fancyhdr,array}
\usepackage[arrow,matrix,curve]{xy}\SilentMatrices
\def\xyma{\xymatrix@M.7em}
\def\xymas{\xymatrix@M.1em}
\newcommand{\be}{\begin{equation}}
\newcommand{\ee}{\end{equation}}
\newcommand{\ba}{\begin{eqnarray}}
\newcommand{\ea}{\end{eqnarray}}

\def\nn{\nonumber}

\def\b{\beta}

\def\l{\lambda}

\def\m{\mu}

\def\IR{\relax{\rm I\kern-.18em R}}
\def\ZZ{\relax{\hbox{\cmss Z\kern-.4em Z}}}

\begin{document}
\begin{titlepage}
\begin{flushright}
DFPD/02/TH/31\\
ULB--TH--02/38\\
hep-th/0212131\\
\end{flushright}

\begin{center} {\Large{\bf Massless spin-two field S-duality}}
\end{center}
\vspace{.3cm}
\begin{center} {\large Xavier Bekaert$^\clubsuit$ and Nicolas Boulanger$^\star$\footnote{``Chercheur
F.R.I.A.'', Belgium} }
\end{center}

\begin{center}{\sl
$^\clubsuit$ Dipartimento di Fisica, Universit\`a degli Studi di Padova\\ Via
F. Marzolo 8, 35131 Padova, Italy}\end{center}
\begin{center}{\sl
$^\star$ Facult\'e des Sciences, Universit\'e Libre de Bruxelles,\\
Campus Plaine C.P. 231, B--1050 Bruxelles, Belgium }\end{center}

\vspace{.3cm}

\begin{abstract}
We present a review of the homological algebra tools involved in
the standard de Rham theory and their subsequent generalizations
relevant for the understanding of free massless higher spin gauge
structure. M-theory arguments suggest the existence of an
extension of (Abelian) S-duality symmetry for non-Abelian gauge
theories, like the four dimensional Yang-Mills or Einstein
theories. Some no-go theorems prove that this extension, if it
exists, should fall outside the scope of local perturbative field
theory.

{\it Proceedings of the RTN-Workshop ``The quantum structure of
spacetime and the geometric nature of fundamental interactions'',
Leuven, September 2002}
\end{abstract}

\vspace{.1cm}

\end{titlepage}
\section{Free higher spin fields}
%

Despite several decades of study, the problem of constructing
covariant consistent interactions for higher-spin fields (i.e.
spin $S>2$) is still only partially solved, and has turned out to
be among the most intriguing and challenging problems in field
theory. From the higher-spin field perspective, it is a
commonplace to {\it{a posteriori}} view string theory as another
attempt to find such a consistent interacting theory.

The birth of higher-spin field quest can be traced back to the
early thirties with the pioneering work of Majorana
\cite{Majorana} which, surprisingly enough, attracted almost no
attention during three decades. In the year 1936, several authors
(among whom Dirac, Klein and Proca) independently addressed the
issue of field equations for particles with spin higher than one.
Still, the ``official" seminal paper on the subject dates from
1939 when Fierz and Pauli \cite{Fierz:1939} wrote the free
covariant actions for spin three-half and two. The
Rarita-Schwinger spin-tensor description of spin-$\frac32$ field
followed two years later \cite{Rarita} (it was obtained earlier by
Tamm in the thirties but he did not published his result). It was
then natural to pursue the systematic study of free fields
propagating in a four-dimensional Minkowski spacetime and, in the
sixties, Chang \cite{Chang} constructed Lagrangians for massive
free fields with spin $S$ between $\frac32$ and $4$. This work was
generalized later on by Singh and Hagen for arbitrary spins
\cite{Singh}. In the bosonic case, this formulation requires the
introduction of $S-1$ auxiliary traceless symmetric tensor fields
of decreasing rank. By taking the massless limit, Fang and
Fronsdal obtained in the late seventies the covariant Lagrangians
for massless fields of any spin \cite{Fronsdal:1978}, elegantly
described by a single gauge field subject to a double
tracelessness condition. It was followed by an alternative
approach to free massless higher spin fields, the so called
``gauge approach" introduced by Vasilev \cite{Vasiliev:1980},
which uses geometrical objects generalizing vielbeins and spin
connection. This approach turned out to be promising for switching
on consistent interactions \cite{Fradkin:1986}.

The four-dimensional case is a very special one, because all the
irreps of the little group $SO(2)$ are completely symmetric.
Furthermore the spin is, strictly speaking, only well-defined in
four dimensions. In higher dimensions, other irreps are possible
and (in the massive case) indeed appear in the spectrum of string
theory. The $D\geq 4$ generalization of the previous results is
direct for the two most simple irreps of the Lorentz group: (i)
completely symmetric fields \cite{deWit} where the $D$-dimensional
massive field actions were obtained from the $(D+1)$-dimensional
massless actions by standard dimensional reduction techniques
\cite{Aragone}; (ii) totally antisymmetric fields, the covariant
formulation of which being easily obtained using differential
forms. As we advocate in this paper, it is in fact possible to
understand the gauge structure of tensor fields in an arbitrary
irreps of $GL(D,\mathbb R)$, already investigated in
\cite{Labastida:1986}, using a systematic unified mathematical
framework generalizing de Rham's theory
\cite{Dubois-Violette:1999,Dubois-Violette:2001,Bekaert:2002}.
Note the recent work \cite{Kirsch:2001ya} based on irreps of
$SL(D,\mathbb R)$, which is then broken to its Lorentz subgroup.
The surge of interest on string field theory in the mid-eighties
prompted the construction of covariant action principles for free
fields in exotic representations of the Lorentz group $SO(D-1,1)$
\cite{Curt,Aula,Labastida:1989} (see \cite{Tsulaia} for a recent
approach).

In a recent work on higher-spin gauge fields, Francia and Sagnotti
\cite{Francia:2002} discovered that forgoing the orthodox locality
requirement allows to relax the tracelessness conditions of the
Fang-Fronsdal formulation. In this non-local approach, the
higher-spin tensor gauge parameters are not constrained to be
irreducible under $SO(D-1,1)$ but only under $GL(D,\mathbb R)$.
For arbitrary spin $S$, gauge invariant field equations were
elegantly written \cite{Francia:2002} in terms of the curvature
tensor introduced by de Wit and Freedman \cite{deWit}. All known
different formulations of free massless higher spin fields exhibit
new features with respect to spin $S\leq 2$ fields (e.g.
tracelessness conditions, auxiliary fields, non-locality, higher
order kinetic operators, etc). This unavoidable novelty of higher
spins is rooted in the fact that the curvature tensor contains $S$
derivatives. Therefore any local higher spin field equation built
from it should be of higher order derivative. This property arises
naturally when formulated in the mathematical framework of
generalized complex developed by Dubois-Violette and Henneaux (for
an extensive review, take a look at \cite{Dubois-Violette:2000})
as was shown for completely symmetric irreps in
\cite{Dubois-Violette:1999,Dubois-Violette:2001} and subsequently
extended to arbitrary irreps in \cite{Bekaert:2002}. We spend most
part of this paper to review this mathematical machinery, that
physicists are not expected to be familiar with.

\section{Algebraic background}

We start with basic definitions from homological algebra. Let
$\{V_i\}_{i\in G}$ be a family of vector spaces indexed by an
Abelian group $G$. The direct sum $V=\oplus_i V_i$ is called the
$G$-{\bf graded space} associated with the family $\{V_i\}_{i\in
G}$. A {\bf differential complex} is defined to be an $\mathbb
N$-graded space $V=\oplus_{i\in \mathbb N} V_i$ with a nilpotent
endomorphism $d$ of degree one, i.e. there is a chain of linear
transformations
\ba\ldots\stackrel{d}{\longrightarrow}V_{i-1}\stackrel{d}{\longrightarrow}
V_i\stackrel{d}{\longrightarrow}V_{i+1}\stackrel{d}{\longrightarrow}\ldots\nn\ea
such that $d^2=0$. A famous example of such structure is the de
Rham complex for which the vector space is the set of differential
forms graded by the form degree. The role of the nilpotent
operator is played by the exterior derivative
$d=dx^\m\partial_\m$. One can now define the quotient
$H(d)\equiv\frac{\mbox{Ker}_d}{\mbox{Im}_d}$ called the {\bf
cohomology} of $d$. This space inherits the grading of $V$. The
elements of $H(d)$ are called ({\bf co}-){\bf cycles}. Elements of
$\mbox{Im}_d$ are said to be trivial or {\bf exact} (co)-cycles.

A straightforward generalization of the previous definitions is to
consider a more complicated grading. More specifically, one takes
$\mathbb N^S$ as Abelian group ($S\geq 2$). A {\bf multicomplex}
is defined to be an $\mathbb N^S$-graded space
$V=\oplus_{(i_1,\ldots,i_S)\in\mathbb N^S} V_{(i_1,\ldots,i_S)}$
with $S$ nilpotent endomorphisms $d_j$ ($1\leq j\leq S$) such that
\ba d_j V_{(i_1,\ldots,i_j,\ldots,n_S)}\subset
V_{(i_1,\ldots,i_j+1,\ldots,i_S)}.\nn\ea A concrete realization of
this definition is the space of {\bf multiforms} whose elements
are sums of products of the generators $d_j x^\mu$ with smooth
functions. Multiforms were recently used in free tensor gauge
theories \cite{Dubois-Violette:2001,Bekaert:2002,deMedeiros:2002}.
For later purposes, we take each of the $d_jx^\mu$,
$j\in\{1,\ldots,S\}$, to be commuting with all the others :
$d_ix^\mu d_jx^\nu=d_jx^\nu d_ix^\mu$, $i\neq j$. When $i=j$, we
demand the natural anticommutativity $d_ix^\mu d_ix^\nu=-d_ix^\nu
d_ix^\mu$. The nilpotent operators $d_j\equiv d_j
x^\mu\partial_\mu$ generalize the exterior differential of the de
Rham complex.

An other possibility is to consider higher orders of nilpotency
which allow a richer set of possibilities for cohomology. An {\bf
$N$-complex} is defined as a graded space $V=\oplus_i V_i$
equipped with an endomorphism $d$ of degree $1$ that is nilpotent
of order $N$: $d^N=0$ \cite{Dubois-Violette:2000}. The {\bf
generalized cohomology} of the $N$-complex $V$ is the family of
$N-1$ graded spaces $H_{(k)}(d)$ with $1\leq k \leq N-1$ defined
by $H_{(k)}(d)=\mbox{Ker}(d^k)/\mbox{Im}(d^{N-k})$, i.e.
$H_{(k)}(d)=\oplus_iH^i_{(k)}(d)$ where
\[
H^i_{(k)}(d)=\left\{ \phi\in V^i\,\,|\,\, d^k\phi=0,\,\,
\phi\sim\phi+d^{N-k}\b,\,\,\b\in V^{p+k-N} \right\}.
\]

The multicomplex structure can be related to the $N$-complex
structure in the following construction which plays a crucial role
in the gauge structure of mixed symmetry type gauge fields. The
idea is rather simple: to connect the two definitions one should
build a $\mathbb N$-grading from the $\mathbb N^S$-grading of the
multicomplex $(V,d_j)$. A simple choice is to consider the {\bf
total grading} defined by the sum $i\equiv \sum_{j=1}^S i_j$. We
introduce the operator $d\equiv \sum_{j=1}^S d_j$ which possesses
the nice property of being of total degree one. Two convenient
cases arise:
\begin{itemize}
  \item $[d_i,d_j]_+=0$ : Usually the nilpotent operators $d_j$ are taken to be
anticommuting and therefore $d$ is nilpotent. This case is rather
standard in homological perturbation theory.
  \item $[d_i,d_j]_-=0$ when $i\neq j$ and ${d_i}^2=0$ :
From our present perspective, commuting $d_j$'s are indeed quite
interesting because, in that case, $d$ is in general nilpotent of
order $S+1$ and the space $V$ is endowed with a $(S+1)$-complex
structure (as far as the multiforms are concerned, this convenient
choice was already made in \cite{deMedeiros:2002}, while the
authors of \cite{Dubois-Violette:2001} chose the previous one).
Indeed, every term in the expansion of $d^{S+1}$ contains at least
one of the $d_j$ twice.
\end{itemize}
In the second case, the {\bf total cohomology group}
$H^{(i_1,\ldots,i_S)}_{(k)}(d)$ is the generalized cohomology
group whose elements $\phi\in V_{(i_1,\ldots,i_S)}$ satisfy the
set of cocycle conditions \be \prod_{j\in J}\,\,\,d_j\,\,\,\phi
\,\,\,=\,\,\, 0\,, \quad\quad \forall J \subset \{1,2, \dots, S\}
\; \,\vert\, \, \# J = k\,,\,\label{cocycle}\ee with the
equivalence relation \be\phi\,\,\,\sim\,\,\,\phi\,\,\,+
\sum_{\begin{array}{c}
J \subset \{1,2, \dots, S\}\\
\#J = S - k +1
\end{array}}d^J\b_J\,,\label{equivalence}\ee
as can be seen easily by decomposing the cocycle condition
$d^k\phi=0$ and the equivalence relation
$\phi\sim\phi+d^{S-k+1}\b$ in $\mathbb N^S$ degree. To our
knowledge, the first appearance of such conditions was in
\cite{Olver:1987} where the total cohomology was christened
``hypercohomology". A sketchy definition of the total cohomology,
underlying its main features, is as the quotient \ba
H^{(i_1,\ldots,i_S)}_{(k)}(d)=\frac{\bigcap\,\mbox{Ker}\,d^k}{\sum\,\mbox{Im}
\,d^{S-k+1}}\,.\nn \ea As a first example of generalization of
Poincar\'e lemma, we mention that the authors of
\cite{Dubois-Violette:2001} proved that (it is a particular case
of their theorem 2 \cite{Dubois-Violette:2001})
$H^{(i_1,\ldots,i_S)}_{(k)}(d)$ is trivial for $i_j\leq D$ in the
subspace of multiforms that vanish at the origin together with all
their $k$ successive derivatives.

\section{Tensor irreps of the general linear group}

Tensor irreps of $GL(D,\mathbb R)$ are characterized by Young
diagrams. A {\bf Young diagram} $Y$ is a diagram which consists of
a finite number $S>0$ of columns of identical squares (referred to
as the {\bf cells}) of finite non-increasing lengths $i_1\geq
i_2\geq \ldots\geq i_S\geq 0$. The total number of cells of the
Young diagram $Y$ is denoted by $|Y|=\sum_{j=1}^S i_j$. A Young
diagram $(i_1,\ldots,i_S)$ is {\bf well-included} into
$(j_1,\ldots,j_S)$ iff $i_k\leq j_k\leq i_k +1$ for all
$k\in\{1,\ldots,S\}$. In other words, the difference between two
Young diagrams does not contain any column greater than a single
box. We denote the well-inclusion as
$(i_1,\ldots,i_S)\Subset(j_1,\ldots,j_S)$.

The set of Young diagrams with at most $S$ columns is denoted by
${\mathbb Y}^S$. We identify any Young diagram $Y$ with its
``coordinates" $(i_1,\ldots,i_S)$. The set ${\mathbb Y}^2$ of
Young diagrams can be pictured as the following subset of the
lattice ${\mathbb N}^2$ : \ba
\xymas{&&&&\ldots\\&&&\stackrel{(3,3)}{\bullet}\ar[r]\ar[ur]&\cdots
\\&&\stackrel{(2,2)}{\bullet} \ar[r]\ar[ur]&
\stackrel{(3,2)}{\bullet} \ar[u]\ar[r]\ar[ur]&\cdots \\
& \stackrel{(1,1)}{\bullet}\ar[r]\ar[ur]&
\stackrel{(2,1)}{\bullet} \ar[r]\ar[u]\ar[ur]&
\stackrel{(3,1)}{\bullet}\ar[u]\ar[r]\ar[ur]&\cdots \\
\stackrel{(0,0)}{\bullet} \ar[r]\ar[ur]&
\stackrel{(1,0)}{\bullet}\ar[u]\ar[r]\ar[ur]&
\stackrel{(2,0)}{\bullet}\ar[u]\ar[r]\ar[ur]
&\stackrel{(3,0)}{\bullet}\ar[u]\ar[r]\ar[ur] &\cdots }\nn\ea
where all the arrows corresponds to well-inclusion relation
$\Subset$.

To any Young diagram $Y$ is associated an operator ${\bf Y}$
projecting tensors of rank $|Y|$ on a definite invariant subspace
$V_Y$ under $GL(D,\mathbb R)$. Every cell of $Y$ is related to a
specific index of the tensors in $V_Y$. The projector ${\bf Y}$ is
referred to as the {\bf Young symmetrizer} because the
corresponding tensors possess definite symmetry properties. The
Young diagrams introduce a ${\mathbb N}^S$-graduation of the space
of tensors $V_S=\oplus_{Y\in{\mathbb Y}^S }V_Y$. This vector space
is relevant in spin $S$ theories because it contains all tensors
that may be relevant in the completely symmetric tensor gauge
field theory (e.g. the gauge field, its curvature, their
electric-magnetic duals, the reducibility parameters, ...).

Let $Y$ and $Y'$ be two Young diagrams in ${\mathbb Y}^S$ such
that $Y\Subset Y'$ and $|Y'|=|Y|+1$. One may generalize the
exterior differential by first taking the partial derivative of a
tensor in $V_Y$ and then acting with the Young symmetrizer ${\bf
Y'}$. This provides a set of (at most) S commuting nilpotent maps
${\mathbf Y'}\circ \partial$ corresponding to each available
direction $Y\stackrel{\Subset}{\longrightarrow}Y'$ in ${\mathbb
Y}^S$. Consequently, the space $V_S$ is endowed with the structure
of a multicomplex. Following the construction of the previous
section, the total grading is the total number of cells $|Y|$ and
we introduce the operator $d$ acting in each invariant subspace
$V_Y\subset V_S$ as $d|_{V_Y} \equiv\sum_{Y'\Supset Y} {\mathbf
Y'}\circ\partial$.

The equivalence relations (\ref{equivalence}), with $k=S$,
expresses naturally the gauge freedom of an arbitrary symmetry
type gauge field $\phi\in V_S$, associated to the invariance of
the curvature tensor $R=d^S\phi$ expressed by (\ref{cocycle}). The
higher nilpotency condition $d^{S+1}=0$ readily explains the fact
that the gauge-invariant curvature $R$ contains $S$ derivatives.
Therefore the gauge structure of free irreducible field theory is
encoded in the cohomology group $H^{(i_1,\ldots,i_S)}_{(1)}(d)$.
With this motivation in mind, we provided a theorem in
\cite{Bekaert:2002} which proves the triviality of the generalized
cohomology groups $H^{(i_1,\ldots,i_S)}_{(k)}(d)$ for $1\leq k
\leq S$, $0<i_S$ and $i_1\leq D$, thereby extending the results of
\cite{Dubois-Violette:1999,Dubois-Violette:2001}.

\section{No-go theorems on non-Abelian S-duality}\label{Lingravi}

Dualities are crucial in order to scrutinize non-perturbative
aspects of gauge field and string theories, it is therefore of
relevance to investigate the duality properties of arbitrary
tensor gauge fields. It is well known that the gravity field
equations in four-dimensional spacetime are formally invariant
under a duality rotation. Furthermore, a deep analogy with the
M$5$-brane toroidal compactification scenario for the self-dual
D$3$-brane was conjectured to occur when the six-dimensional (4,0)
superconformal gravity theory is compactified over a $2$-torus
\cite{Hull:2000}. This compactification would lead to a
$SL(2,\mathbb Z)$-duality group for four-dimensional Einstein
gravity, geometrically realized as the modular group of the torus.
However a naive duality rotation is not a true symmetry of gravity
because the covariant derivative involves the gauge field, which
is not inert under a formal duality rotation. We face the same
problem in Yang-Mills theories and some no-go theorems can be
established suggesting that non-Abelian S-duality may fall outside
the scope of local perturbative field theory (see
\cite{Bekaert:2001} for a short review of the spin one case).

At the free level, the electric-magnetic duality scheme works
perfectly for massless tensor fields in arbitrary representations
of $GL(D,\mathbb R)$ \cite{Hull:2001,Bekaert:2002,deMedeiros:2002}
(duality properties of massive spin two-fields were discussed
recently in \cite{Casini:2002}). The main idea is to reinterpret
field equations and Bianchi identites as either
\begin{itemize}
  \item \underline{symmetry conditions} of some (Hodge) dual
$\tilde{R}=\prod_{j\in J}*_jR$ (where $J \subset \{1,2, \dots,
S\}$) of the curvature tensor $R$ (considered as a multiform), or
  \item \underline{cocycle conditions} that allows to derive, from the above
mentioned theorem, the existence of a dual gauge field
$\tilde{\phi}$ such that $\tilde{R}=d^S\tilde{\phi}$.
\end{itemize}

Dualizing the Pauli-Fierz symmetric gauge field $h_{(\mu\nu)}$ in
$D>4$ generates new fields in irreducible representations of
$GL(D,\mathbb R)$, characterized by mixed Young tableaux. There
exists in fact two different dual fields obtained in generic
spacetime dimension $D$.  The first one is obtained by dualizing
on one index only and involves a tensor $T_{\l_1 \l_2 \cdots
\l_{D-3} \m}$ with
\begin{eqnarray}
&& T_{\l_1 \l_2 \cdots \l_{D-3} \m} = T_{[\l_1 \l_2 \cdots
\l_{D-3}] \m}, \\
&&T_{[\l_1 \l_2 \cdots \l_{D-3} \m]} = 0
\end{eqnarray}
where $[ \; \; ]$ denotes antisymmetrization with strength one.
The second one is obtained by dualizing on both indices and is
described by a tensor $C_{\l_1 \cdots \l_{D-3} \m_1 \cdots
\m_{D-3}}$ with Young symmetry type $(D-3,D-3)$ (two columns with
$D-3$ boxes). Although one can write equations of motion for this
theory which are equivalent to the linearized Einstein equations,
these do not seem to follow (when $D>4$) from a Lorentz-invariant
action principle in which the only varied field is $C_{\l_1 \cdots
\l_{D-3}\, \m_1 \cdots \m_{D-3}}$. For this reason, we focus here
on the dual theory based on $T_{\l_1 \l_2 \cdots \l_{D-3} \m}$.
For a recent work where massive fields possessing the
aforementioned symmetries are discussed in flat and (A)dS
background, see \cite{Zinoviev:2002}.

When $D=5$, the dual version of linearized gravity in terms of the
field $T_{\l_1 \l_2  \m}$ was given in  \cite{Curt}, while the
action for arbitrary $D$ can be found in \cite{Aula}. In
\cite{Bekaert:2002uh} we studied the problem of determining all
the consistent, local, smooth interactions that these dual
formulations in terms of $T_{\l_1 \l_2 \cdots \l_{D-3} \m}$ gauge
fields admit. It is well known that the only consistent (local,
smooth) deformation of the Pauli-Fierz theory is - under quite
general and reasonable assumptions - given by the Einstein theory
(see the works \cite{Wald,Boulanger:2000} for systematic
analyses). Because dualization is a non-local process, one does
not expect the Einstein interaction vertex to have a local
counterpart on the dual $T_{\l_1 \l_2 \cdots \l_{D-3} \m}$-side.
This does not {\it{a priori}} preclude the existence of other
local interaction vertices, which would lead to exotic
self-interactions of spin-two particles. The main result of
\cite{Bekaert:2002uh} is, however, that there are no Lorentz
invariant deformations with at most two
derivatives of the field.

\section*{Acknowledgements}
%
X.B. thanks D. Francia and A. Sagnotti for stimulating discussions
while N.B. is grateful to G. Barnich, G. Bonelli and M. Henneaux
for interesting comments. The work is supported in part by the
``Actions de Recherche Concert{\'e}es" of the ``Direction de la
Recherche Scientifique - Communaut\'e Francaise de Belgique",
IISN-Belgium (convention 4.4505.86), a ``P\^ole d'Attraction
Interuniversitaire" (Belgium) and the European Commission RTN
programme HPRN-CT-2000-00131, in which N.B. is associated with
K.U. Leuven.
%


\end{document}